\documentclass[a4paper,12pt]{article}
\usepackage{a4}
\usepackage{epsfig}
\usepackage{latexsym, amssymb} 
\usepackage{amsmath}
\usepackage{psfrag}
\usepackage{bm}
\usepackage{feynmf}
\usepackage{rotating}
\usepackage{hyperref}
\usepackage[utf8]{inputenc}
\hypersetup{
    unicode=false,          
    pdftoolbar=true,        
    pdfmenubar=true,        
    pdffitwindow=false,     
    pdfstartview={FitH},    
    pdftitle={My title},    
    pdfauthor={Author},     
    pdfsubject={Subject},   
    pdfcreator={Creator},   
    pdfproducer={Producer}, 
    pdfkeywords={keyword1} {key2} {key3}, 
    pdfnewwindow=true,      
    colorlinks=true,        
    linkcolor=red,          
    citecolor=blue,         
    filecolor=magenta,      
    urlcolor=green,         
    linktocpage=true
}

\def\Mpl{\mathrm{M}_{_{\mathrm{Pl}}}}

\def\beq{\begin{equation}}
\def\eeq{\end{equation}}
\def\br{\begin{eqnarray}}
\def\er{\end{eqnarray}}
\def\benu{\begin{enumerate}}
\def\efnu{\end{enumerate}}

\def\l{\left}
\def\r{\right}
\def\cR{{\cal R}}
\def\d{{\rm d}}
\def\f{\frac}

\def\cR{{\cal R}}

\def\Mp{\rm M_{_{\rm Pl}}}

\begin{document}
\title{Changes in the halo formation rates due to\\ 
features in the primordial spectrum}
\author{Dhiraj Kumar Hazra\footnote{Current address: Asia Pacific Center for Theoretical Physics, Pohang, 
Gyeongbuk 790-784, Korea.
\vskip 5pt
E-mail:~dhiraj@apctp.org}\\
Harish-Chandra Research Institute Chhatnag Road,\\ Jhunsi, Allahabad~211019, India}
\maketitle
\begin{abstract}
Features in the primordial scalar power spectrum provide a possible roadway 
to describe the outliers at the low multipoles in the WMAP data.
Apart from the CMB angular power spectrum, these features can also alter the 
matter power spectrum and, thereby, the formation of the large scale structure. 
Carrying out a complete numerical analysis, we investigate the effects of 
primordial features on the formation rates of the halos. 
We consider a few different inflationary models that lead to features in the
scalar power spectrum and an improved fit to the CMB data, and analyze the 
corresponding imprints on the formation of halos.
Performing a Markov Chain Monte Carlo analysis with the WMAP seven year data 
and the SDSS halo power spectrum from LRG DR7 for the models of our interest, 
we arrive at the parameter space of the models allowed by the data.
We illustrate that, inflationary potentials, such as the quadratic potential 
with sinusoidal modulations and the axion monodromy model, which generate 
certain repeated, oscillatory features in the inflationary perturbation 
spectrum, do not induce a substantial difference in the number density of halos 
at their best fit values, when compared with, say, a nearly scale invariant 
spectrum as is generated by the standard quadratic potential. 
However, we find that the number density and the formation rates of halos 
change by about $13$--$22\%$ for halo masses ranging over $10^4$--$10^{14}\, 
M_{\odot}$, for potential parameters that lie within $2$-$\sigma$ around the 
best fit values arrived at from the aforesaid joint constraints. 
We briefly discuss the implications of our results.
\end{abstract}

\newpage
\tableofcontents

\section{Introduction}

Eversince its original proposal more than three decades ago, inflation 
has continued to be the most efficient scenario to overcome the horizon 
and the flatness problems associated with the conventional, hot big bang 
model. 
Additionally, inflation also provides an efficient mechanism for sowing the 
seeds of the density perturbations in the early universe~\cite{texts,reviews}. 
Over the last few years, observations of the Cosmic Microwave Background 
(CMB) by missions such as the Wilkinson's Microwave Anisotropy Probe 
(WMAP)~\cite{wmap-5,wmap-7} and the Atacama Cosmology Telescope 
(ACT)~\cite{act-2010} have provided us with observational bounds on the 
parameters that describe the power spectra of the primordial perturbations. 
Currently, apart from the strong constraints on the primordial scalar amplitude 
and the spectral index, there also exists an upper bound on the tensor-to-scalar 
ratio. 
Clearly, the primordial power spectra generated by inflation must be 
consistent with these observational constraints.

\par

The simplest mechanism to achieve inflation consists of a canonical 
scalar field (referred to as the inflaton) which rolls down slowly 
towards the minima of its potential. 
The perturbations originating from such a scalar field produces an
almost scale independent power spectrum~\cite{texts,reviews}. 
Such a primordial spectrum, along with a suitable choice of parameters 
to describe the background $\Lambda$CDM model, is found to lead to a CMB 
angular power spectrum which fits the available observational data quite 
well.
In fact, despite the constantly improving bounds, one finds that there 
exist many inflationary models that remain consistent with the 
observations. 

\par 

However, there exist a few points (notably near the multipoles of $\ell=2$, 
$22$ and $40$) in the WMAP data~\cite{wmap-5,wmap-7}, which lie outside the 
cosmic variance associated with the best fit theoretical curve corresponding 
to the CMB angular power spectra generated by slow roll inflation. 
Though such outliers are few in number, demanding a better fit to these 
data points can, obviously, lead to a more favored model of inflation. 
Several attempts have been made to construct inflationary potentials that 
lead to wiggles in the primordial power spectrum (which, in turn, cause 
corresponding oscillations in the CMB angular power spectrum), thereby 
resulting in an improved fit to the data when compared to the featureless 
power spectra produced by slow roll 
inflation~\cite{l2240,joy-2008-2009,pi,hazra-2010}. 
The wiggles can be generated if the inflaton goes through a step or a bump
in the potential. 
The presence of a step or a bump leads to a period of fast roll, which 
then leaves its imprints on the slow roll parameters and the power 
spectrum~\cite{starobinsky-1992,hu-2010-2011}.
Apart from constructing specific models that are more favored by the data, 
there has also been a few efforts towards a model independent reconstruction 
of the primordial scalar power spectrum from the observed CMB angular power 
spectrum~\cite{rc}.
Importantly, such reconstructions too seem to support the presence of 
features in the primordial power spectrum. 
Although the reconstruction efforts suggest that the features in the scalar 
power spectrum are typically localized over certain scales, it has been found 
that, oscillatory potentials which generate features on all scales also provide 
a much better fit to the CMB data than the nearly scale invariant power 
spectrum~\cite{pahud-2009,flauger-2010,aich-2011}. 
It may be worth mentioning here that inflationary models leading to features 
also gain importance due to the fact that they can generate large levels of 
non-Gaussianity as is possibly indicated by the WMAP data (in this context, 
see, for example, Refs.~\cite{ng,ng-features} and the references therein).

\par

Evidently, any significant change in the primordial power spectrum will 
correspondingly modify the matter power spectrum evaluated today. 
Hence, the features in the scalar perturbation spectrum that we mentioned
above will affect the matter power spectrum which, in turn, will leave its 
signatures on the formation rates of the dark matter halos.
The effects of steps or oscillations in the inflaton potential on the 
predicted number of halos was recently investigated based on the perturbation 
spectrum evaluated using the slow roll approximation~\cite{rodrigues-2010}.
However, the slow roll parameters can often turn large in such inflationary 
potentials and, as a result, the power spectrum evaluated in the slow roll 
approximation can differ considerably from the actual power spectrum.
Moreover, with the ever increasing quality of the large scale structure
observations, in particular, the halo power spectrum constructed using 
the Luminous Red Galaxies (LRG) from the seventh data release (DR7) of 
the Sloan Digital Sky Survey (SDSS)~\cite{sdss,sdss-data}, 
it would be a worthwhile exercise to actually compare the models with the 
available data to arrive at additional constraints (i.e. apart from those
obtained from the CMB data) on the primordial features.

\par

In this paper, we utilize a recently developed and accurate Fortran~90 
code to {\it exactly}\/ evaluate the inflationary perturbation spectrum, 
and thereby the corresponding matter power spectrum and the number of halos 
formed.
We compute the number densities and the formation rates of halos in a couple 
of inflationary models that are known to lead to certain features in the power 
spectrum and also an improved fit to the CMB data.
Our goal will be to estimate the maximum possible change in the number 
density and the formation rate of halos in these models, when compared 
with, say, inflation driven by the simplest quadratic potential. 
In order to arrive at the parameter space of interest, we shall compare 
the models with the CMB as well as the large scale structure data. 
Specifically, we shall make use of the WMAP seven year (WMAP-$7$) 
data~\cite{wmap-7} and the LRG halo power spectrum data from SDSS 
DR7~\cite{sdss-data}.
We find that the power spectra with features corresponding to 
the best fit values of the inflationary parameters do not typically lead 
to substantial deviations in the formation rates of halos.
However, we find that, in certain models that we consider, the potential 
parameters that lie within $2$-$\sigma$ of the best fit values obtained 
from the joint constraints of the WMAP and the SDSS data can lead to a 
$13$--$22\%$ change in the number of halos formed for halo masses ranging 
over $10^4$--$10^{14}\, M_{\odot}$.

\par

This paper is organized as follows.
In the following section, we shall quickly describe the inflationary 
potentials of interest, which lead to specific features in the 
primordial scalar power spectrum. 
We shall also outline the method that we adopt to compare the models with 
the data. 
In Section~\ref{sec:formalism}, we shall outline the formalism to 
arrive at the matter power spectra and the halo formation rates 
from the inflationary power spectra. 
In Section~\ref{sec:nm}, we shall provide a few essential details concerning 
the numerical procedures that we follow.
We shall discuss the results in Section~\ref{sec:results}, and we shall close 
with a few concluding remarks regarding the wider implications of our results 
in Section~\ref{sec:discussion}.

\par

Note that we shall assume the background cosmological model to be the 
standard, spatially flat, $\Lambda$CDM model.
 

\section{Primordial spectra with features}

In this section, we shall quickly sketch a few essential points 
concerning the scalar power spectrum generated during inflation.
We shall also discuss the inflationary models of our interest and 
the scalar power spectra produced by them.
Further, we shall outline the methods that we adopt to compare the
models with the data.


\subsection{Essentials}

We shall focus on the simplest case of inflation driven by the canonical 
scalar field, say, $\phi$.
In a spatially flat Friedmann universe, given a potential $V(\phi)$, the 
scalar field satisfies the differential equation
\begin{equation}
{\ddot \phi}+3\,H\,\dot{\phi}+V_{\phi}=0,\label{eq:beq}
\end{equation}
where the overdot represents differentiation with respect to the cosmic 
time, $H={\dot a}/a$ is the Hubble parameter, with $a$ being the scale
factor, while $V_{\phi}=\d V/\d \phi$. 
The evolution of the scalar field is usually characterized by a 
hierarchy of the so-called slow roll parameters.
The first slow roll parameter is set to be $\epsilon_1=-\dot{H}/H^2$,
and the higher order slow roll parameters are defined in terms of the
first as follows: $\epsilon_{i+1}=\d\,{\rm ln}\,\epsilon_{i}/\d\,N$, 
where $i\ge1$ and $N$ denotes the number of e-folds. 

\par
 
The scalar perturbations induced by the quantum fluctuations of the
inflaton can be described by the curvature perturbation, say, $\cR$,
whose Fourier modes satisfy the following differential equation:
\begin{equation}
\cR_{k}''+2\, \frac{z'}{z}\, \cR_{k}'
+k^{2}\, \cR_{k} = 0,\label{eq:req}
\end{equation}
where the overprime denotes differentiation with respect to the 
conformal time coordinate, and $z=a\,\dot{\phi}/H$. 
Upon solving this differential equation with suitable initial conditions
for modes of cosmological interest, one arrives at the scalar power 
spectrum (see Section~\ref{sec:nm} for the necessary details), defined as
\begin{equation}
{\cal P}_{_{\rm S}}(k)
=\f{k^3}{2\, \pi^{2}}\; \vert \cR_{k}\vert^{2},\label{eq:pps}
\end{equation}
with $\cR_{k}$ being evaluated at sufficiently late times.
Smooth potentials, in general, permit slowly rolling fields (i.e.
wherein the parameters $\epsilon_i$ are much smaller than unity), 
and such backgrounds generate nearly scale invariant power spectra.
For instance, the archetypical quadratic potential, viz. $V(\phi)=
m^{2}\,\phi^{2}/2$, leads to the required COBE amplitude (for a suitable
value of $m$)  and an almost scale invariant spectrum with a scalar 
spectral index of $n_{_{\rm S}} \simeq 0.97$, in strong conformity 
with the recent CMB data~\cite{wmap-5,wmap-7}.


\subsection{Models of interest}
  
In contrast to smooth potentials, potentials that contain either
non-trivial forms or sharp changes in their slopes lead to 
deviations from slow roll.
Departures from slow roll affect the amplitude of the modes that 
leave the Hubble radius during this period resulting in specific 
features in the inflationary scalar power spectrum, with the shape
being determined by the type of deviation from slow roll (in this 
context, see, for instance, Ref.~\cite{jain-2007}).
Interestingly, certain features in the scalar power spectrum are 
known to allow an improved fit to the CMB data than the conventional 
nearly scale invariant spectrum, as is generated by slow roll 
inflation~\cite{l2240,joy-2008-2009,pi,hazra-2010,pahud-2009,flauger-2010,aich-2011}.

\par 

\begin{figure}[!htb]
\psfrag{N}[0][1][1.5]{$ N$} 
\psfrag{eps1}[0][1][1.5]{$\epsilon_1$}
\psfrag{eps2}[0][1][1.5]{$\epsilon_2$}
\resizebox{230pt}{175pt}{\includegraphics{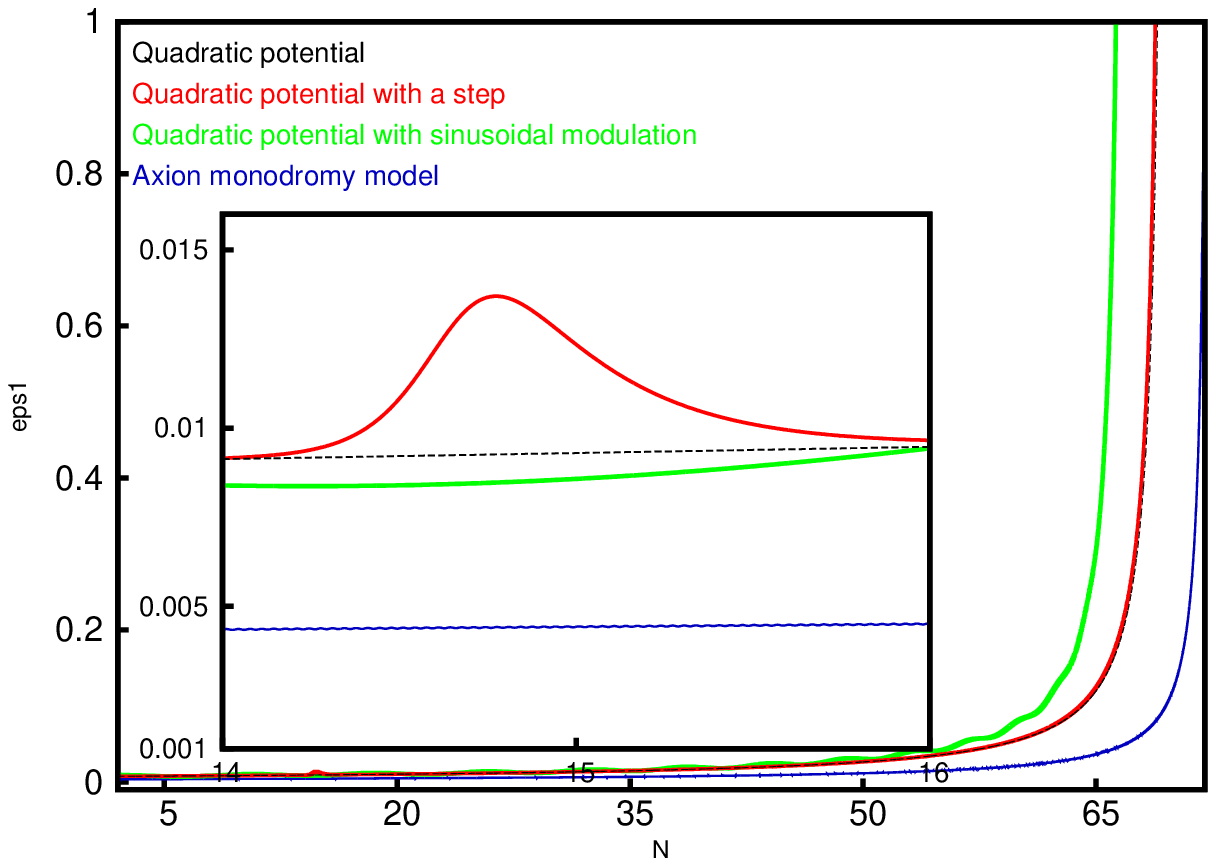}}
\resizebox{230pt}{175pt}{\includegraphics{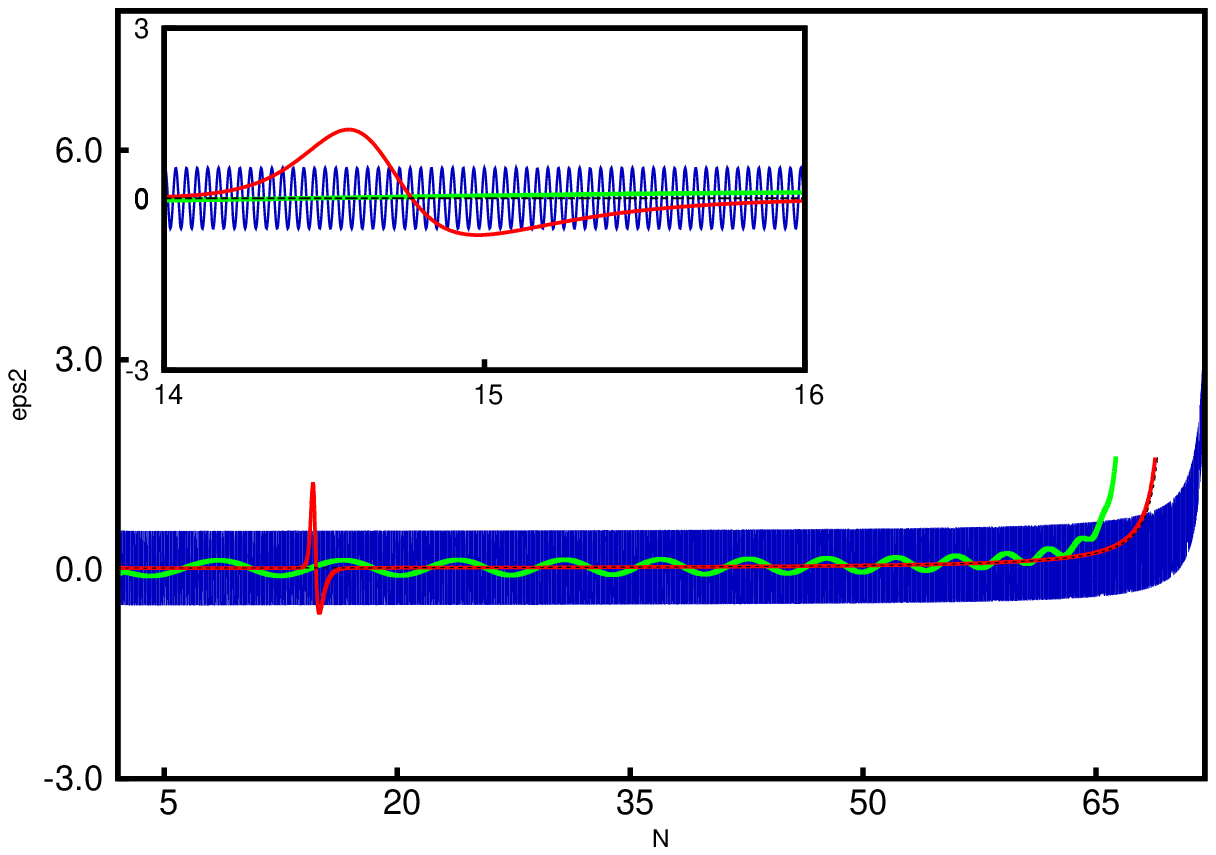}}
\caption{\footnotesize\label{fig:slow-roll-parameters} Evolution of the
first two slow roll parameters $\epsilon_1$ and $\epsilon_2$ have been 
plotted as a function of the number of e-folds $N$ for the quadratic 
potential without and with the step (the black and the red curves, 
respectively), the quadratic potential superimposed by sinusoidal modulations 
(the green curve) and the axion monodromy model (the blue curve). 
The curves have been plotted for potential parameters that lead to the
best fit to the recent WMAP and SDSS data. 
In the quadratic potential without the step, as is well known, the field 
continues to roll slowly until the end of inflation, whereas, when the step 
is introduced, it briefly deviates from slow roll around the time the field 
crosses the step. 
In the case of the potentials with oscillations, for the best fit values, 
the axion monodromy model leads to strong departures from slow roll, with 
$\epsilon_2$ turning large repeatedly, right till the termination of 
inflation.
The insets provide a closer view of the behavior of the slow roll parameters 
over a smaller range of e-folds.}
\end{figure}

In this work, we shall consider two types of inflationary models
that have been shown to fit the data better than the power law 
primordial spectrum.
The first type of model that we shall consider contains a step, which 
is typically introduced in the conventional quadratic potential at a
certain location, say, $\phi_0$, as follows~\cite{l2240,hazra-2010}:
\begin{equation}
V(\phi) = \frac{1}{2}\,m^2\,\phi^2\,  
\l[1 + \alpha\,\tanh\l(\frac{\phi-\phi_{0}}{\Delta\phi}\r) \r].
\label{eq:qpstep}
\end{equation} 
Evidently, $\alpha$ and $\Delta\phi$ denote the strength and the width 
of the step, respectively.
The field experiences a short period of fast roll as it crosses $\phi_{0}$,
leading to a brief burst of oscillations in the scalar power spectrum.
A Markov Chain Monte Carlo (MCMC) sampling of the parameter space of the 
above potential points to the fact that the presence of the step aids in 
fitting the outliers in the WMAP data near the multipoles of $\ell=22$ 
and $40$.
In fact, at the expense of the three additional parameters, viz. $\phi_0$,
$\alpha$ and $\Delta\phi$, the model results in an improvement in the 
effective least squares parameter $\chi_{\rm eff}^{2}$ by about $9$, when 
compared with the quadratic potential without the step.
In Figure~\ref{fig:slow-roll-parameters}, we have plotted the behavior 
of the first two slow roll parameters in the model. 
 
\par

Over the last few years, two types of oscillatory inflationary 
potentials have drawn a considerable amount 
attention~\cite{pahud-2009,flauger-2010,aich-2011,ng-features}.
The first of these consists of the above-mentioned quadratic 
potential that is modulated by sinusoidal oscillations as 
follows:
\begin{equation}
V(\phi) = \frac{1}{2}\,m^2\,\phi^2  \l[1 + \alpha\,
\sin\l(\frac{\phi}{\beta}+\delta\r) \r].\label{eq:qpsin}
\end{equation}    
The second is the so-called axion monodromy model that has its 
motivations in string theory, and is described by the potential
\begin{equation}
V(\phi) = \lambda\, \l[\phi
+\alpha~\Mpl\,\cos\l(\frac{\phi}{\beta}+\delta\r)\r].\label{eq:amm} 
\end{equation}
In both these cases, the slow roll parameters oscillate and, in
fact, continue to exhibit such a behavior until the termination 
of inflation (cf. Figure~\ref{fig:slow-roll-parameters}). 
All modes of interest are affected by these oscillations, which 
leave a repeated pattern that extends over all scales in the power 
spectrum. 
We shall illustrate the power spectra resulting in these models later 
in Section~\ref{sec:results} (see Figure~\ref{fig:sps-all}), wherein we 
shall present the results of our analysis.  
We would like to mention here that recent analysis indicate that such 
power spectra lead to an improved fit to the WMAP as well as the small 
scale ACT data.   
In the case of the quadratic potential modulated by sinusoidal oscillations, 
one finds that the least squares parameter $\chi_{\rm eff}^{2}$ reduces by 
about unity, whereas the monodromy model is found to lead to a much better 
fit to the data with an improvement in $\chi_{\rm eff}^{2}$ of up to $13$,
when compared with a more conventional, nearly scale invariant 
spectrum~\cite{flauger-2010,aich-2011}. 

\subsection{Comparison with the WMAP and the SDSS data}

We shall compare the models with the CMB as well as the large scale 
structure data. 
We have worked with the WMAP-$7$ data~\cite{wmap-7} and the halo power 
spectrum data arrived at from the LRG in SDSS DR7~\cite{sdss-data}. 
We have made use of the large scale structure data to ensure that the 
parameter values we eventually work with to obtain the formation rate 
of the halos are consistent with the observed matter power spectrum. 

\par

We have made use of the publicly available codes, viz. the cosmological 
Boltzmann code CAMB~\cite{camb,lewis-2000} and the Monte Carlo code
COSMOMC~\cite{cosmomc,lewis-2002}, to compute the CMB angular and the 
matter power spectra, and compare them with the data, respectively. 
As we had mentioned, we shall assume the background model to be the 
spatially flat $\Lambda$CDM and we shall work with the priors on the 
background parameters as listed in Table~\ref{tab:priors-bp}.
\begin{table}[!htb]
\begin{center}
\begin{tabular}{|c|c|c|}
\hline
Background & Lower & Upper\\ 
parameter & limit & limit\\
\hline
$\Omega_{\rm b}\, h^2$ & $0.005$ & $0.1$\\
\hline
$\Omega_{\rm c}\, h^2$ & $0.01$ & $0.99$\\
\hline
$\theta$ & $0.5$ & $10.0$\\
\hline
$\tau$ & $0.01$  & $0.8$\\
\hline
\end{tabular}
\caption{\footnotesize\label{tab:priors-bp}The priors on the four parameters 
that describe the background, spatially flat, $\Lambda$CDM model. 
The quantities $\Omega_{\rm b}\, h^2$ and $\Omega_{\rm c}\, h^2$ describe the 
baryon and CDM densities (with $h$ being related to the Hubble parameter), 
$\theta$ is the ratio of the sound horizon to the angular diameter distance at 
decoupling, while $\tau$ is the optical depth to reionization.
We should mention that we keep the same priors on the background parameters for 
all the COSMOMC analysis in this paper.}
\end{center}
\end{table}

\par

We should note here that the step model~(\ref{eq:qpstep}), the 
quadratic potential with the sinusoidal modulation~(\ref{eq:qpsin}) 
and the axion monodromy model~(\ref{eq:amm}), all require four 
parameters to describe the potential completely. 
As far as the priors on these parameters are concerned, for the case of the 
quadratic potential with the step, we have worked with the same priors as we
had worked with before~\cite{hazra-2010}. 
In the case of the quadratic potential with superimposed sinusoidal 
modulations and the axion monodromy model, for the primary parameters $m$ 
and $\lambda$, we work with the same priors that we had considered in a 
recent analysis wherein we had compared the models with the CMB 
data~\cite{aich-2011}.
However, while comparing with the WMAP as well as the SDSS data, we have 
widened the priors of the parameters $\alpha$ and $\beta$ for both the 
oscillatory potentials as listed in Table~\ref{tab:priors-pp}. 
We should add that we have allowed the phase parameter $\delta$ to vary 
from $-\pi$ to $\pi$ as before.
\begin{table}[!htb]
\begin{center}
\begin{tabular}{|c|c|c|c|}
\hline
Model & Potential & Lower & Upper\\ 
& parameter & limit & limit\\
\hline
Quadratic potential &$\alpha$ & $0$ & $2\times10^{-3}$\\
\cline{2-4}
with sinusoidal modulation& ${\mathrm{ln}}\,  (\beta/\Mpl) $ & $-3.9$ & $0$\\
\hline 
Axion monodromy & $\alpha$ & $0$ & $2\times10^{-4}$\\
\cline{2-4}
model & ${\mathrm{ln}}\, (\beta/\Mpl)$ & $-8$  & $0$\\
\hline
\end{tabular}
\caption{\footnotesize\label{tab:priors-pp}The widened priors on the parameters
$\alpha$ and $\beta$ of the potentials~(\ref{eq:qpsin}) and~(\ref{eq:amm}). 
Note that the priors are considerably wider than the limits in our recent 
analysis~\cite{aich-2011}. 
Also, we have also worked with the logarithmic value of $\beta$ in order to cover
a larger range with a uniform weightage.}
\end{center}
\end{table}

\par

The motivations for these choice of priors are two fold. 
Firstly, while determining the priors on the parameters $m$ and $\lambda$, 
we have ensured that the resulting inflationary power spectra do not differ 
considerably from the nearly scale invariant spectrum for the faster 
convergence of the Markov Chains. 
Secondly, we have chosen the priors on $\alpha$ and $\beta$ such that the 
scalar field does not get trapped by the oscillations in the potential. 

\par

We should mention that, unlike in our earlier efforts~\cite{hazra-2010,aich-2011}, 
we have not taken the effects of the tensor perturbations into account, as the 
corresponding effects are negligible. 
Moreover, in the case of potentials with oscillations, which lead to fine features 
in the inflationary scalar power spectrum, we actually need to modify CAMB in order 
to ensure that the CMB angular power spectrum is evaluated at every multipole and 
compare them with the data~\cite{flauger-2010,aich-2011,huang}. 
But, we have not implemented this point here since we are only interested in the 
marginalized probabilities of the potential parameters. 
These probabilities shall indicate the extent to which deviations from a nearly 
scale invariant spectrum is allowed by the data, and the corresponding effects 
on the formation of dark matter halos. 
We should point out that we have not taken into account the non-linear effects 
on the matter power spectrum~\cite{nonlinear}, but have included the SZ effect 
and the effects due to gravitational lensing in our analysis. 
Finally, we shall set the Gelman and Rubin parameter $\vert R-1\vert$ to be $0.03$ 
for convergence in all the cases.

 
\section{From the primordial spectrum to the formation rate of
halos}\label{sec:formalism}

In this section, we shall quickly outline the standard formalism to arrive 
at the formation rate of halos from the primordial power spectrum.


\subsection{The matter power spectrum}

Given a primordial power spectrum, say, ${\cal P}_{_{\rm S}}(k)$ 
[cf.~Eq.~(\ref{eq:pps})], the matter power spectrum at the 
redshift~$z$ is usually written as (see, for example, 
Refs.~\cite{mo-2010,takada-2006})
\begin{eqnarray}
P_{_{\rm M}}(k,z) 
&=& \l(\frac{2}{5\,\Omega_{\rm m}}\r)^{2}\, \l(\frac{k}{a_0\,H_0}\r)^4\,
\l(\f{2\,\pi^2}{k^3}\r)\,{\cal P}_{_{\rm S}}(k)\,T^{2}(k)\,D_{+}^{2}(z),
\label{eq:mps}  
\end{eqnarray}
a quantity that can be conveniently expressed in units of ${\rm Mpc}^3$ 
for the modes of cosmological interest. 
In the above expression, $T(k)$ is the CDM transfer function, the quantity 
$D_{+}(z)$ denotes the linear growth factor of the total matter perturbation,
$\Omega_{\rm m}$ is the non-relativistic density parameter, while $a_0$ and 
$H_0$ denote the scale factor and the Hubble parameter today.   

\par 

If we define $D_{+}(a)=g(a)/a$, then, one finds that, in the spatially 
flat $\Lambda$CDM model, the quantity $g$ satisfies the differential 
equation~\cite{wmap-5,gf}
\begin{equation}
\frac{\d^2g}{\d\ln a^2} 
+ \frac{1}{2}\,\l[5+3\,\Omega_{\Lambda}(a)\r]\,\frac{\d g}{\d \ln a}
+3\,\Omega_{\rm eff}(a)\,g=0 ,\label{eq:gf}
\end{equation}
where $\Omega_{\rm eff}(a)= \Omega_{\Lambda}\, H_0^2/H^2$, with 
$\Omega_{\Lambda}$ denoting the dimensionless density parameter
associated with the cosmological constant today.
In this work, we shall solve the above differential equation with 
suitable initial conditions to obtain the growth factor $D_{+}(a)$.  
Utilizing CAMB~\cite{camb,lewis-2000} to determine the corresponding 
transfer function $T(k)$ and, upon using the primordial spectrum obtained 
numerically in the inflationary model of interest, we shall eventually 
arrive at the matter power spectrum (see Section~\ref{sec:nm} on numerical 
methods for further details). 


\subsection{Mass functions and the halo formation rates}

To arrive at the formation rate of dark matter halos, we shall first
require the number density of collapsed halos with mass in the range 
of $M$ and $M+\delta M$ in a comoving volume element. 
This number density, say, $n(M)$, is defined in terms of the root mean 
square fluctuation in mass $\sigma$ through the so-called mass function 
$f({\sigma})$ as follows~\cite{jenkins-2001}:
\begin{equation}
\frac{\d n}{\d\ln M}
=-\frac{2\,\rho_{\rm m}}{M}\l(\frac{\d\ln\sigma(R)}{\d
\ln M}\r)\,f(\sigma),\label{eq:dndlnm}
\end{equation}
where $\rho_{\rm m}$ is the mean density of non-relativistic matter in 
the universe. 
Following the convention (see, for example, Ref.~\cite{mo-2010}), we 
shall define the root mean square fluctuation in mass at the scale $R$ 
to be 
\beq
\sigma^2(R)
=\int_{0}^{\infty}\d\ln k\; {\cal P}_{_{\rm M}}(k)\;{\widetilde W}^2(k,R),
\label{eq:sigma}
\eeq
where ${\cal P}_{_{\rm M}}(k)\equiv k^3\,P_{_{\rm M}}(k)/(2\,\pi^2)$ denotes 
the {\it dimensionless}\/ matter power spectrum, while ${\widetilde W}(k,R)$ is 
the Fourier transform of the window function $W(x,R)$ that is introduced to 
smooth out the density perturbation.
We shall work with the commonly used spherical top hat window function, whose 
Fourier transform is given by
\begin{equation}
\widetilde{W}(k,R)
=\widetilde{W}(k\,R)
=3~\frac{\sin\, (k\,R)-k\,R\,\cos\, (k\,R)}{(k\,R)^3},
\end{equation}
corresponding to the volume $V(R)=4\,\pi\,R^3/3$.
Note that the halo mass $M$ within the window of radius $R$ is given by $M(R)
=\rho_{\rm m}\,V(R)$. 
 
\par

We shall make use of the Sheth-Tormen mass function to evaluate the 
number density of halos~\cite{sheth-2001-2002}.
In contrast to the more conventional Press-Schechter mass 
function~\cite{press-1974}, it has been found that the Sheth-Tormen 
mass function fits the data from the $N$-body simulations better. 
Actually, the Sheth-Tormen mass function is a generalization of the original
Press-Schechter formalism for spherical collapse to the case of ellipsoidal 
collapse. 
The Sheth-Tormen mass function is defined in terms of two additional parameters 
$b$ and $p$ (when compared to the Press-Schechter case)  
as follows:
\beq  
f(\sigma)
=A\, \sqrt{\frac{b\,\nu}{2\,\pi}}\; \l[1+(b\,\nu)^{-p}\r]\;
\exp-\l(b\,\nu/2\r),\label{eq:stmf}
\eeq
where $\nu=\l(\delta_{\rm c}/\sigma\r)^2$, with $\delta_{\rm c}=1.686$ being
the threshold linear overdensity for collapse. 
The Press-Schechter mass function corresponds to $A=1/2$, $b=1$ and $p=0$. 
However, upon comparing with the $N$-body simulation data, the best fit values
for $b$ and $p$ are found out to be $0.707$ and $0.3$, respectively. 
The value of $A$ can then be arrived at from the normalization condition
on $f(\nu)$, viz. that the integral of $f(\nu)/\nu$ over all $\nu$ is unity, 
which leads to $A=0.3222$.

\par

The number density of halos associated with the above Sheth-Tormen mass 
function is then given by
\begin{eqnarray}
\frac{\d n}{\d \ln M}
&=&-\frac{A\, \rho_{\rm m}}{M}\,
\sqrt{\frac{2\,b\,\nu}{\pi}}\,
\l[1+(b\,\nu)^{-p}\r]\,\l(\frac{\d\ln \sigma}{\d\ln M}\r)\,
\exp-\l(b\,\nu/2\r).\label{eq:dndlnmst}
\end{eqnarray}
The corresponding formation rates of the halos can be easily obtained to 
be~\cite{hfr}
\begin{eqnarray} 
R(M,z)
&=&-\frac{\d D_{+}(z)}{\d z}\;\frac{\d z}{\d t}\; \frac{1}{D_{+}(z)}\, 
\l[\frac{2\,p}{1+(b\,\nu)^{-p}}-b\,\nu\r]\,
\frac{\d n}{\d M}.\label{eq:fr}
\end{eqnarray}
Note that the quantity $\d D_{+}/\d z$ proves to be 
negative, since the growth factor decreases as the redshift increases.
As a result, it is known that the above formation rate of halos can 
become negative for some mass scales (i.e. when $2\,p/[1+(b\,
\nu)^{-p}]>b\,\nu$), which in practice can not occur.
Therefore, to avoid this issue and simultaneously illustrate the effects 
of features, we shall only plot the ratio of the formation rates in the 
inflationary models leading to features and the conventional, smooth, 
quadratic potential.


\section{Details of the numerical methods}\label{sec:nm}

The slow roll approximation allows the background as well as the 
perturbations to be evaluated analytically during inflation.
As we pointed out before, in the earlier work~\cite{rodrigues-2010}, 
it was the slow roll approximation that was made use of in order to 
arrive at the inflationary perturbation spectrum ${\cal P}_{_{\rm S}}(k)$.
However, we find that, in the inflationary models of interest, the slow 
roll parameters turn sufficiently large (cf. Figure~\ref{fig:slow-roll-parameters}) 
implying a breakdown of the slow roll approximation.  
As a result, we resort to numerical methods to compute the primordial 
perturbation spectrum. 
In fact, we utilize a Fortran~90 code that has been recently developed by us, 
which makes use of a Bulirsch-Stoer algorithm along with an adaptive step size 
control routine~\cite{press-1996} to accurately and efficiently solve the 
equations governing the background and the perturbations, in order to arrive at 
the inflationary scalar power spectrum (see Refs.~\cite{hazra-2010,aich-2011} 
for further details in this regard).

\par
 
Having computed the primordial power spectrum, we arrive at the matter 
power spectrum using the transfer function and the growth factor. 
As we had remarked earlier, we obtain the transfer function from CAMB, 
and we evaluate the growth factor by solving the differential 
equation~(\ref{eq:gf}). 
It should be mentioned here that the initial conditions are chosen 
such that $g$ is a constant and equal to unity in the early matter 
dominated epoch, i.e. at a sufficiently high redshift of, say, $z
\simeq 30$~\cite{wmap-5,komatsu-code}. 

\par

After having obtained the matter power spectrum, we calculate the 
variance $\sigma(R)$ using equation~(\ref{eq:sigma}).
The integral can be evaluated numerically with the simplest of algorithms, 
provided the power spectrum proves to be smooth and devoid of any features.
In contrast, when these exist features such as repeated oscillations, 
certain care is required, and we have made use of an adaptive integration 
routine to compute the integral involved~\cite{gq}. 
We have carried out integral from a suitably small mode (such as $k=10^{-5}\,
{\rm Mpc}^{-1}$) up to a mode where the window function cuts off the integrand. 
Finally, we obtain the quantity $\d\ln\sigma/\d\ln M$ by numerical 
differentiation.  
We should stress here that, keeping in mind the 
presence of oscillations in the power spectra, we have computed 
$\sigma$ and $\d\ln\sigma/\d\ln M$ with care and high accuracy.
We should also add that we have cross checked our result by fitting 
the numerical values of $\sigma(R)$ to the Chebyshev polynomials and 
calculating the corresponding derivative from the polynomial (in this 
context, see Ref.~\cite{komatsu-code}). 
 

\section{Results}\label{sec:results}

In this section, we shall present the results of our comparison of the 
models of our interest with the CMB and the large scale structure data. 
We shall also discuss the effects of primordial spectra with features 
on the formation of halos.


\subsection{Joint constraints from the WMAP and the SDSS data}

In Table~\ref{tab:bestfit} below, we have tabulated the best fit values 
of the background and the potential parameters obtained from the MCMC 
analysis using the WMAP-$7$ and the SDSS LRG DR7 data. 
We have also listed the effective least squared parameter $\chi^{2}_{\rm eff}$ 
in each of the cases. 
For the case of the quadratic potential with and without the step, we 
have arrived at results similar to what we have obtained in an earlier 
work~\cite{hazra-2010}.
Also, as one would expect, we find that the background parameters 
are better constrained with the inclusion of the additional SDSS 
data~\cite{finelli-2010,benetti-2012}.
Moreover, it is obvious from Table~\ref{tab:bestfit} that the axion monodromy 
model does not lead to the same extent of improvement in the fit as has been 
obtained before (in this context, see Refs.~\cite{flauger-2010,aich-2011}).
This arises due to the fact that, unlike in the earlier analysis, we have not 
evaluated the CMB angular power at each multipole, but have worked with the 
inbuilt effective sampling and interpolation routine in CAMB. 
However, we should stress that this does not affect our conclusions since our 
focus here lies on the maximum change in the formation of halos. 
Therefore, we are more interested in the allowed regions of the parameter space 
of rather in arriving at the precise best fit point. 
Also, importantly, as we shall discuss in the following subsection, for violent 
oscillations in the primordial power spectrum (when one requires computing the 
CMB angular power spectrum at each multipole explicitly and accurately), 
the percentage change in the number density of halos proves to be negligible 
in the observable mass bins.
\begin{table}[!htb]
\label{sine-param}
\begin{center}
\begin{tabular}{|c|c|c|c|c|}
\hline
Model & Quadratic & Quadratic~$+$~step & Quadratic~$+$~sine & Axion 
monodromy\\
\hline
$\Omega_{\mathrm{b}}\, h^2$ & 0.0222 & 0.0221 & 0.0216&0.0225\\
\hline
$\Omega_{\mathrm{c}}\, h^2$ & 0.1162 & 0.1159 & 0.1168&0.1154\\
\hline
$\theta$ & 1.038 & 1.039 & 1.036& 1.039\\
\hline
$\tau$ & 0.0824 & 0.0875 &0.0836 &0.0856\\ 
\hline
${\mathrm{ln}}\, \l(10^{10}\, A\r)$ & -0.6545 & -0.6406 &-0.6448 & 0.9649\\
\hline
$\alpha$ & - & $1.61\times 10^{-3}$ &$6.35\times 10^{-5}$ &$4.4\times 10^{-5}$\\
\hline
$\phi_0/\Mpl$ & - & 14.664 & -&-\\
\hline
$\Delta\phi/\Mpl$ & - & $3.22\times 10^{-3}$ &-&-\\ 
\hline
$\mathrm{ln}\, (\beta/\Mpl)$ & - & - & -2.576&-7.61\\
\hline
$\delta$ & - & - & 2.208&-1.178\\
\hline
$\chi_{\mathrm{eff}}^{2}$ & 7515.57 & 7507.3 & 7515.12& 7509.56\\
\hline 
\end{tabular}
\caption{\footnotesize\label{tab:bestfit}The best fit values for the 
background and the potential parameters for the different models of 
interest obtained from the MCMC analysis using the WMAP-$7$ and the 
SDSS LRG DR7 data. 
We should mention here that the parameter $A$ denotes $\lambda/\Mp^3$ 
in the case of axion monodromy model and $m^2/\Mp^2$ in rest of the 
cases. 
As we have discussed before, the quadratic potential with the step 
improves the fit to the outliers in the CMB data around the multipoles 
of $\ell=22~\rm{and}~40$. 
Moreover, as we have pointed out, while the superimposed sinusoidal 
modulation to the quadratic potential does not provide a better fit
to the data when compared to the quadratic potential, the axion 
monodromy model improves the fit to a good extent. 
However, note that, the monodromy model does not improve the fit 
to the data to the same extent that as has been arrived at 
recently~\cite{flauger-2010,aich-2011}.
As we have pointed out in the text, this arises due to the limited 
sampling and interpolation by CAMB over the multipoles of interest.}
\end{center}
\end{table}

\par

\begin{figure}[!htb]
\begin{center}
\psfrag{Likelihood}[0][1][1.5]{Normalized Likelihood}
\psfrag{phi0}[0][1][2]{$\phi_0/\Mpl$}
\psfrag{alpha}[0][1][2.5]{$\alpha$}
\psfrag{lnbeta}[0][1][2.5]{$\ln\, (\beta/\Mpl)$}
\resizebox{160pt}{140pt}{\includegraphics{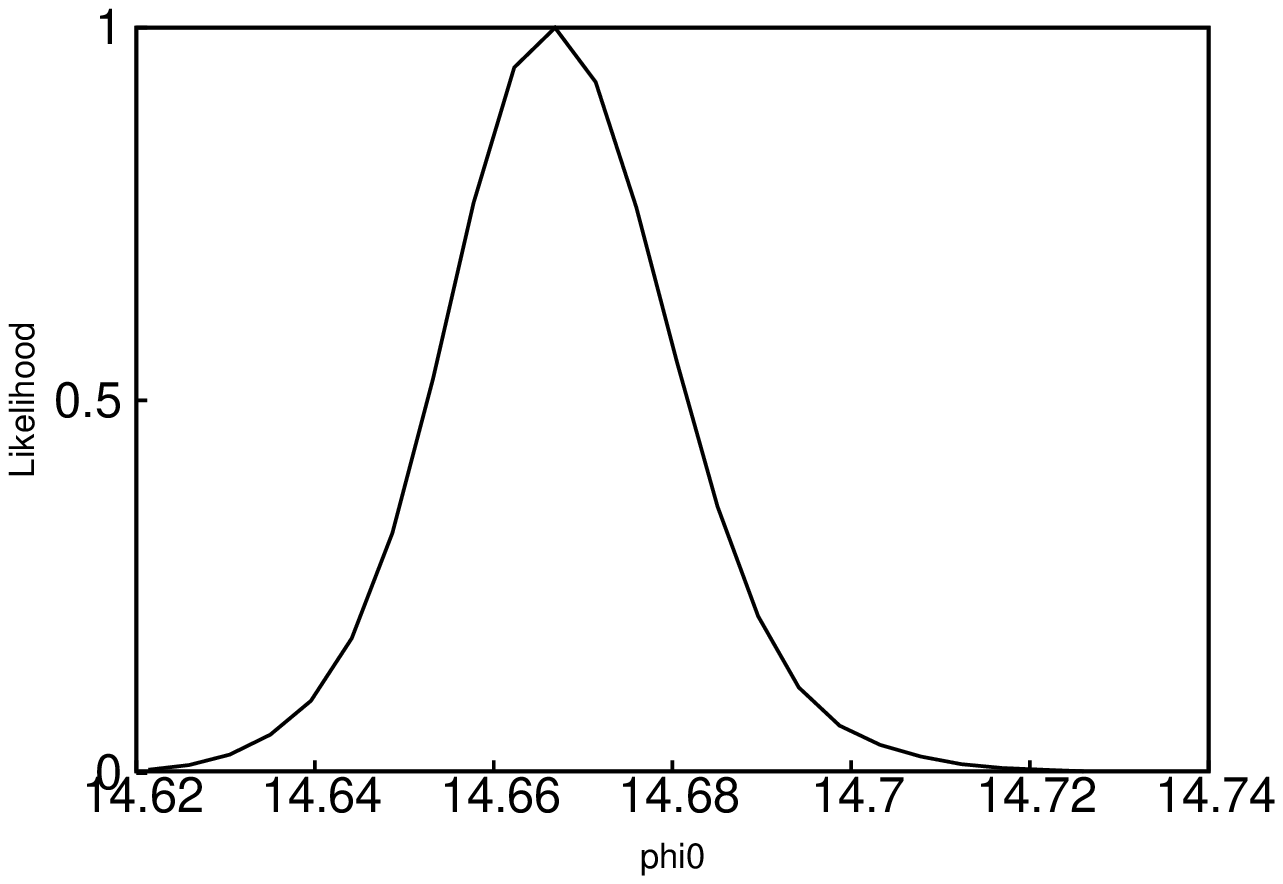}}
\resizebox{140pt}{140pt}{\includegraphics{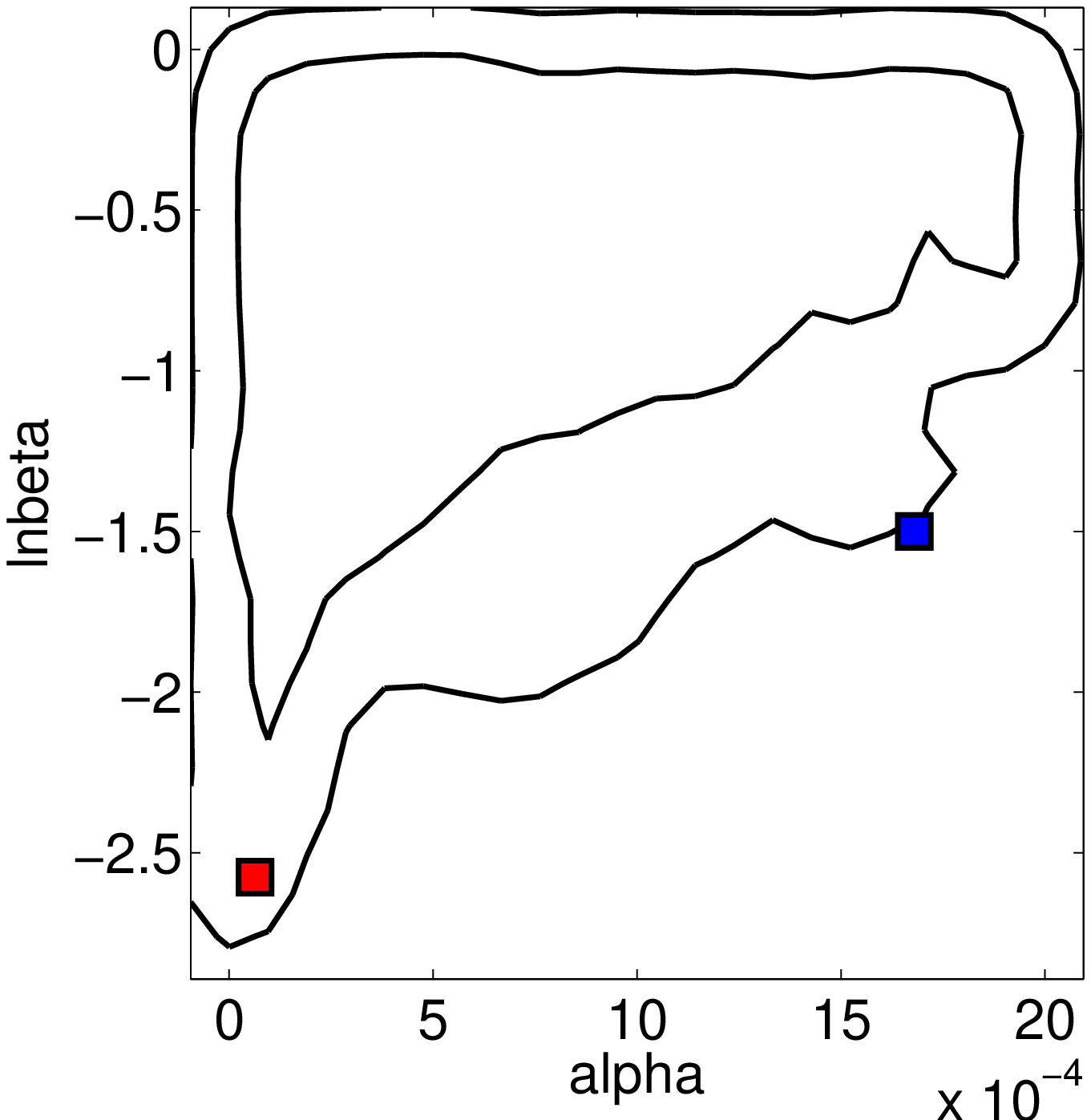}}
\resizebox{140pt}{140pt}{\includegraphics{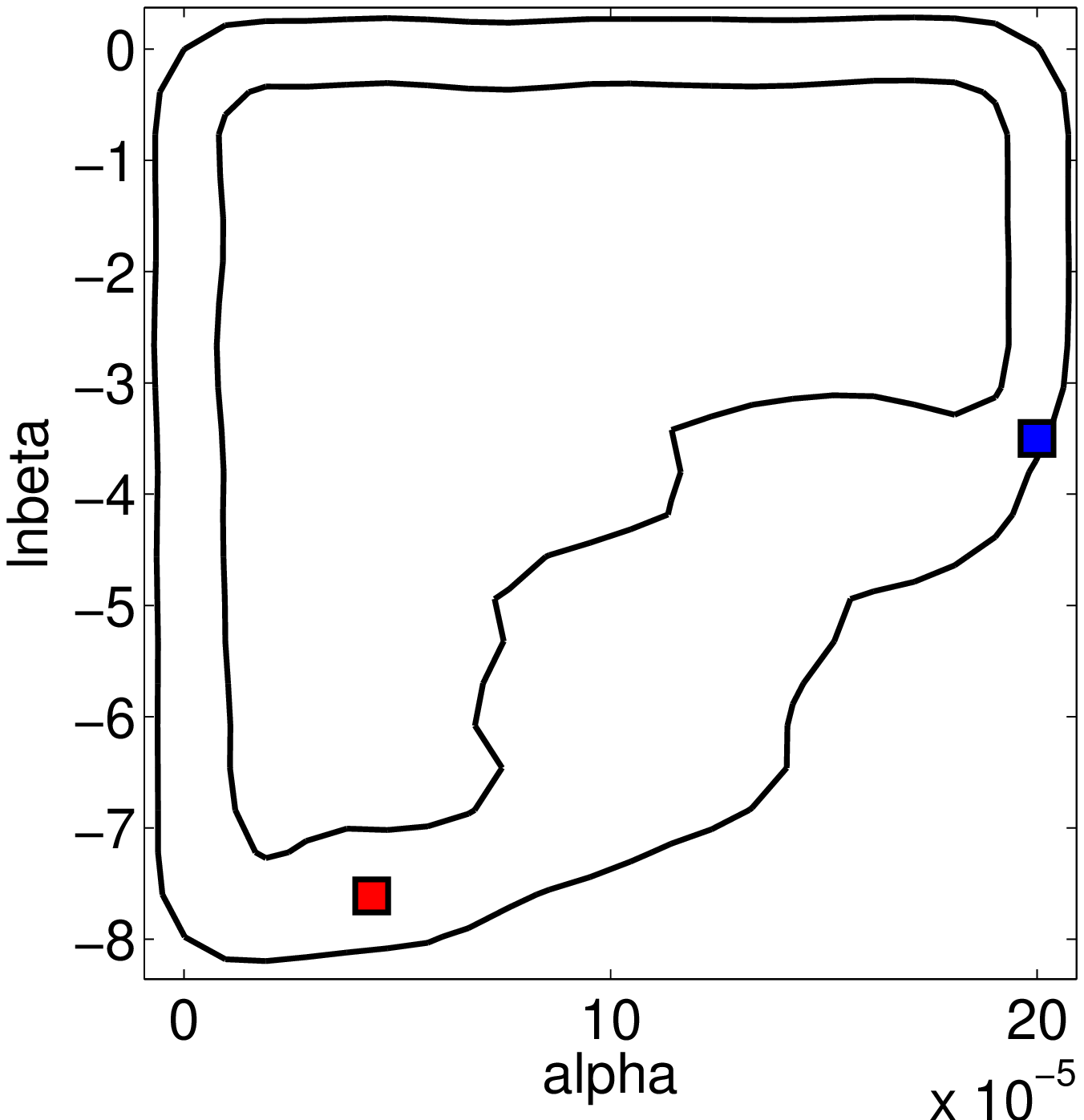}}
\end{center}
\caption{\footnotesize\label{fig:likelihoods}
The one dimensional likelihood on the parameter $\phi_0$ in the case of 
the quadratic potential with a step (on the left) and the two 
dimensional constraints on the parameters $\alpha$ and $\beta$ in the 
cases of the quadratic potential with sinusoidal modulations (in the 
middle) and the axion monodromy model (on the right). 
Note that the location of the step $\phi_0$ is well constrained. 
The inner and the outer curves (in the figure in the middle and the one 
on the right) correspond to the 1-$\sigma$ and the 2-$\sigma$ confidence
contours. 
The red points appearing in the two plots on the right correspond to the 
best fit values of $\alpha$ and $\beta$, while the blue points indicate 
a sample point in the parameter space at and around which the maximum
deviation in number density of halos occurs, when compared to the more
conventional quadratic model (in this context, see Figure~\ref{fig:nh-fbfv}).
We should highlight here the fact that, in the cases of both the
inflationary models with oscillations in the potential (i.e. the
chaotic model with sinusoidal modulation as well as the axion
monodromy model), the best fit values of $\alpha$ and $\ln\,\beta$
lie outside the 1-$\sigma$ marginalized contours.
We find that this occurs because of the reason that the marginalized
one-dimensional probability distributions of these parameters are
highly skewed. This in turn indicates that a good improvement in the fit occurs in a
small region of the parameter space with a low marginalized probability.
}
\end{figure}

In Figure~\ref{fig:likelihoods}, we have illustrated the one dimensional 
likelihood on the parameter $\phi_0$ in the case of the quadratic potential 
with a step, and it is clear that the location of the step is highly 
constrained by data. 
The step affects the number density of halos over only highly localized mass 
scales. 
We have also plotted the marginalized two dimensional constraints on the 
parameters $\alpha$ and $\beta$ for the cases of the two oscillatory potentials. 
It is noteworthy that the constraints are strikingly similar. 
In fact, the roughly triangular shape of the contours can also be understood. 
As the parameter $\beta$ decreases, the resulting oscillations in the potential 
and, therefore, in the inflationary perturbation spectrum turn too frequent, 
and the data constrains the amplitude $\alpha$ to a smaller region.

\par

\begin{figure}[!htb]
\begin{center}
\psfrag{psk}[0][1][1.2]{${\cal P}_{{\rm S}}(k)$} 
\psfrag{kmpc}[0][2][1.2]{$k~{\rm Mpc^{-1}}$}
\resizebox{400pt}{270pt}{\includegraphics{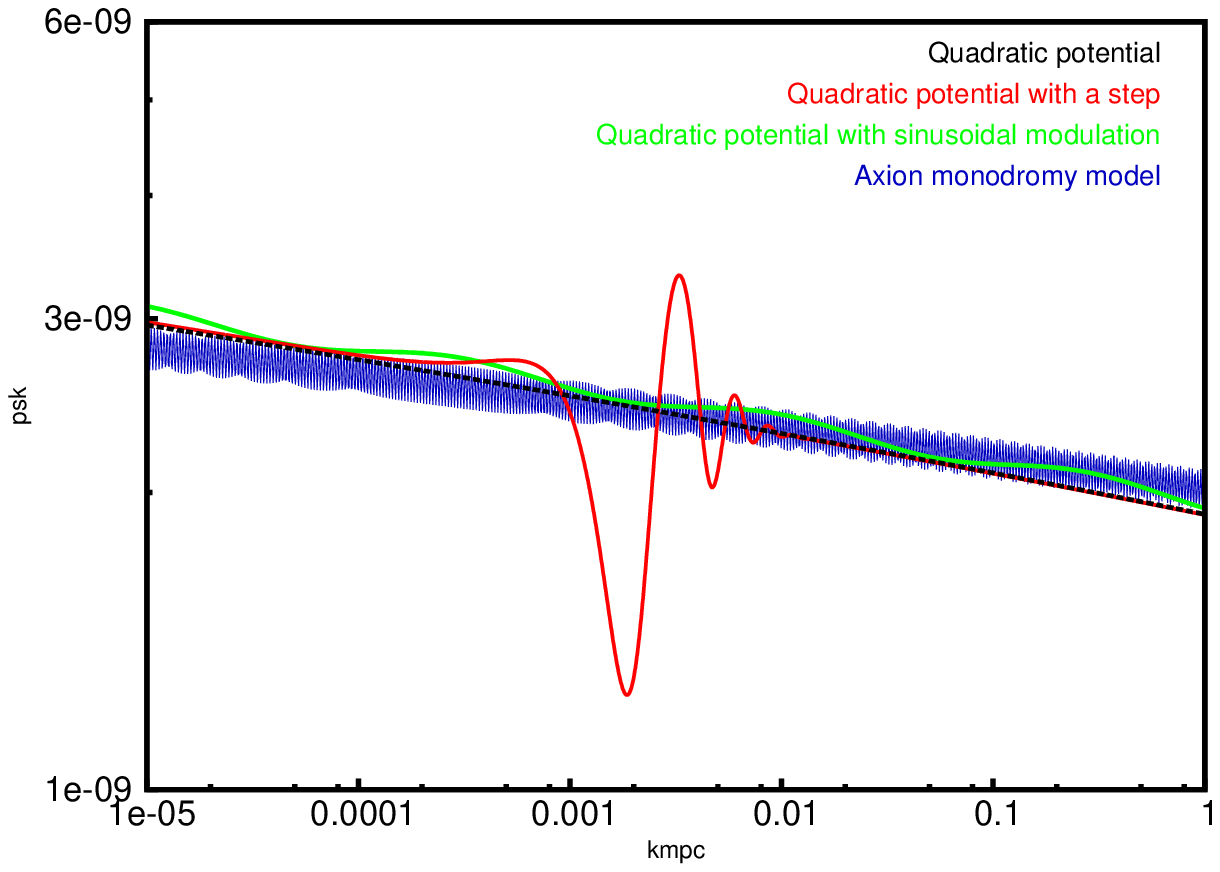}} 
\end{center}
\caption{\footnotesize\label{fig:sps-all} The scalar power spectra 
that arise in the four inflationary models of our interest.  
These spectra correspond to the best fit values for which we 
had plotted the evolution of the first two slow roll parameters 
in Figure~\ref{fig:slow-roll-parameters}.
Note that, we have worked with the same choice of colors to represent 
the results from the different models as in the earlier figure. 
We should emphasize that, while the step model leads to features that 
are localized, the potentials with oscillatory terms lead to modulations 
in the scalar power spectrum that extend over a wide range of scales.}
\end{figure}

\begin{figure}[!htb]
\begin{center}
\psfrag{pmc}[0][1][1.2]{$P_{{\rm M}}(k)~h^{-3}~{\rm Mpc^3} $}
\psfrag{kmpc}[0][1][1.2]{$k~h~{\rm Mpc^{-1}}$}
\resizebox{400pt}{270pt}{\includegraphics{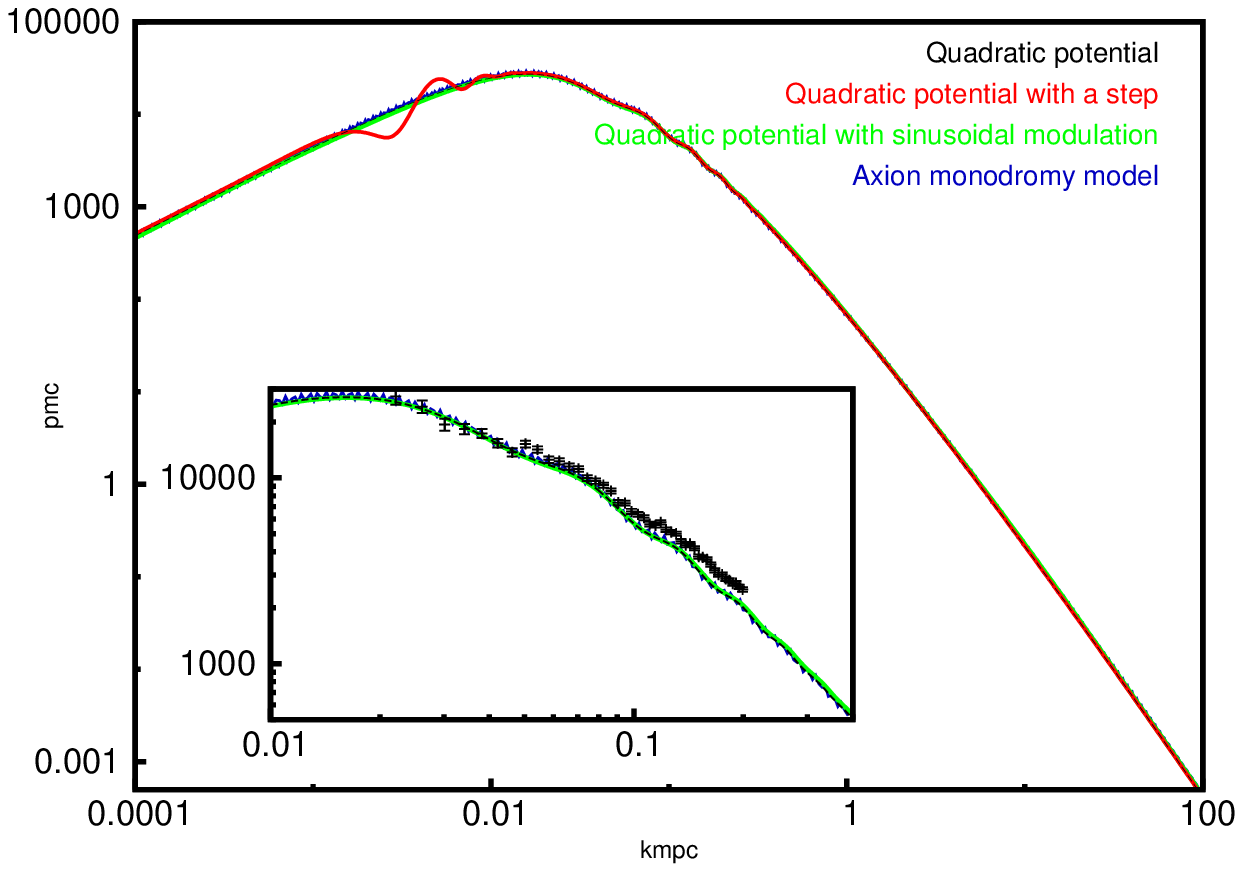}} 
\end{center}
\caption{\footnotesize\label{fig:mps-all} The best fit matter power spectrum 
$P_{_{\rm M}}(k)$ corresponding to the different inflationary scalar power 
spectra we had plotted in the previous figure.
The inset highlights the imprints of the primordial features on the matter 
power spectrum in the domain where the baryon acoustic oscillations also 
play a role. 
The black dots with bars correspond to the power spectrum data from SDSS 
obtained upon combining main galaxies and the LRGs with error bars arrived 
at from the diagonal elements of the corresponding covariance matrix. 
It should be noted that, since we have plotted the linear power spectrum, 
without taking into account the non-linear effects, the theoretical best fit
curves are unable to fit the data well for $k>0.1\, h\,{\rm Mpc^{-1}}$ (in 
this context, see Ref.~\cite{nonlinear}).}
\end{figure}

In Figure~\ref{fig:sps-all}, we have plotted the best fit scalar power 
spectrum for quadratic potential with and without the step and the two 
oscillatory potentials. 
Further below, in Figure~\ref{fig:mps-all}, we have plotted the matter 
power spectrum $P_{M}(k)$ evaluated {\it today}\/ corresponding to the
different inflationary power spectra in the previous figure. 
In the inset of the figure, we have highlighted the baryon acoustic 
oscillations and the halo power spectrum data from SDSS LRG DR7. 
We should add here that the theoretical best fit curves are unable to
fit the data well after $k\sim0.1\,h\,{\rm Mpc}^{-1}$ due to the fact 
that we have not taken the non-linear effects into account in arriving 
at the matter power spectrum.


\subsection{Effects of features on the number density and the formation 
rate of halos}\label{subsec:chfr}

\begin{figure}[!htb]
\begin{center} 
\psfrag{ratrate}[0][1][1.2]{ Change in R}
\psfrag{ratdndlnm}[0][1][1.2]{ Change in $\d n_{\rm ST}/\d \ln M$}
\psfrag{minmsun}[0][1][1.2]{ $M~~in~~M_{\odot}$}
\resizebox{200pt}{250pt}{\includegraphics{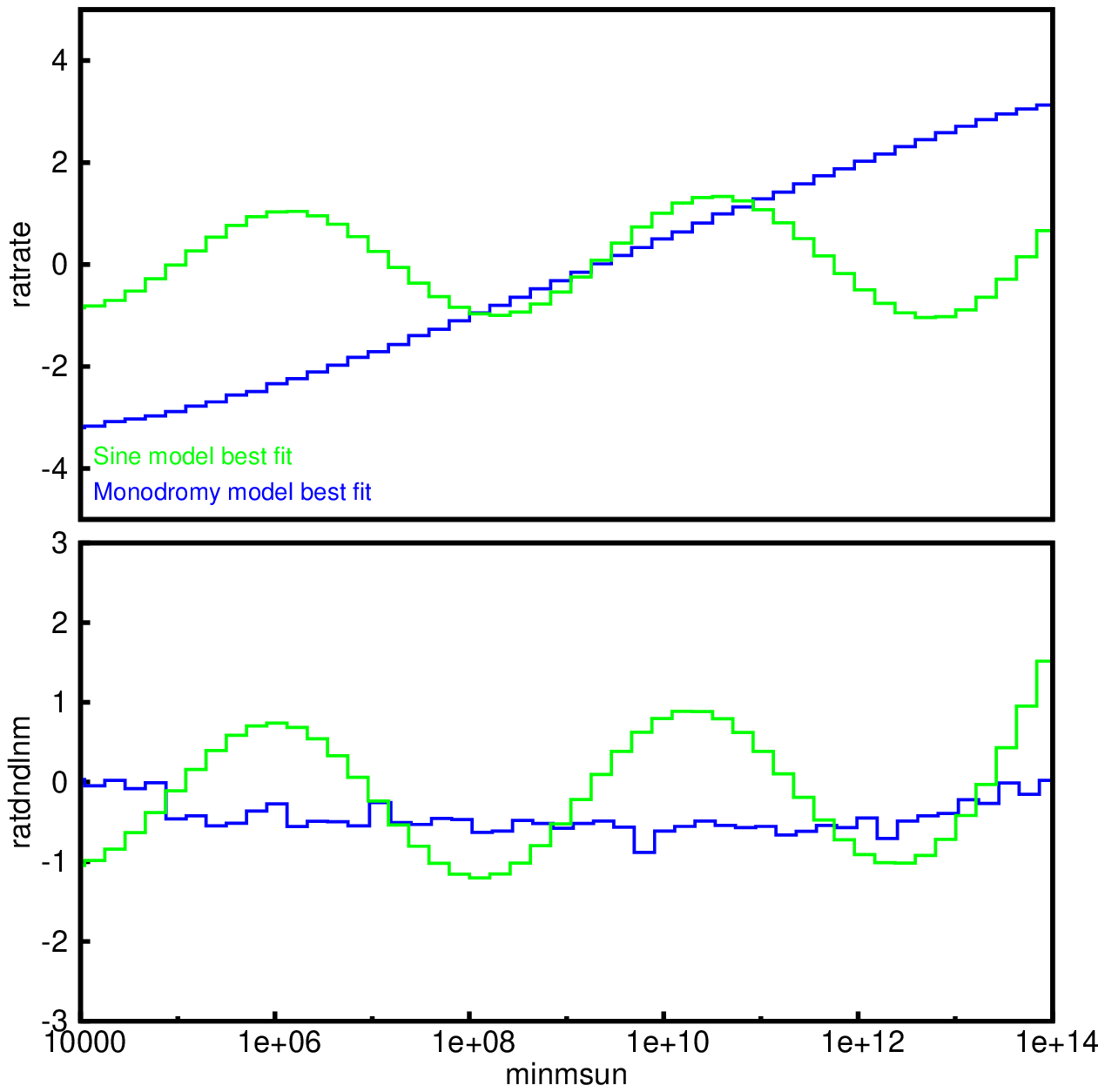}} 
\resizebox{200pt}{250pt}{\includegraphics{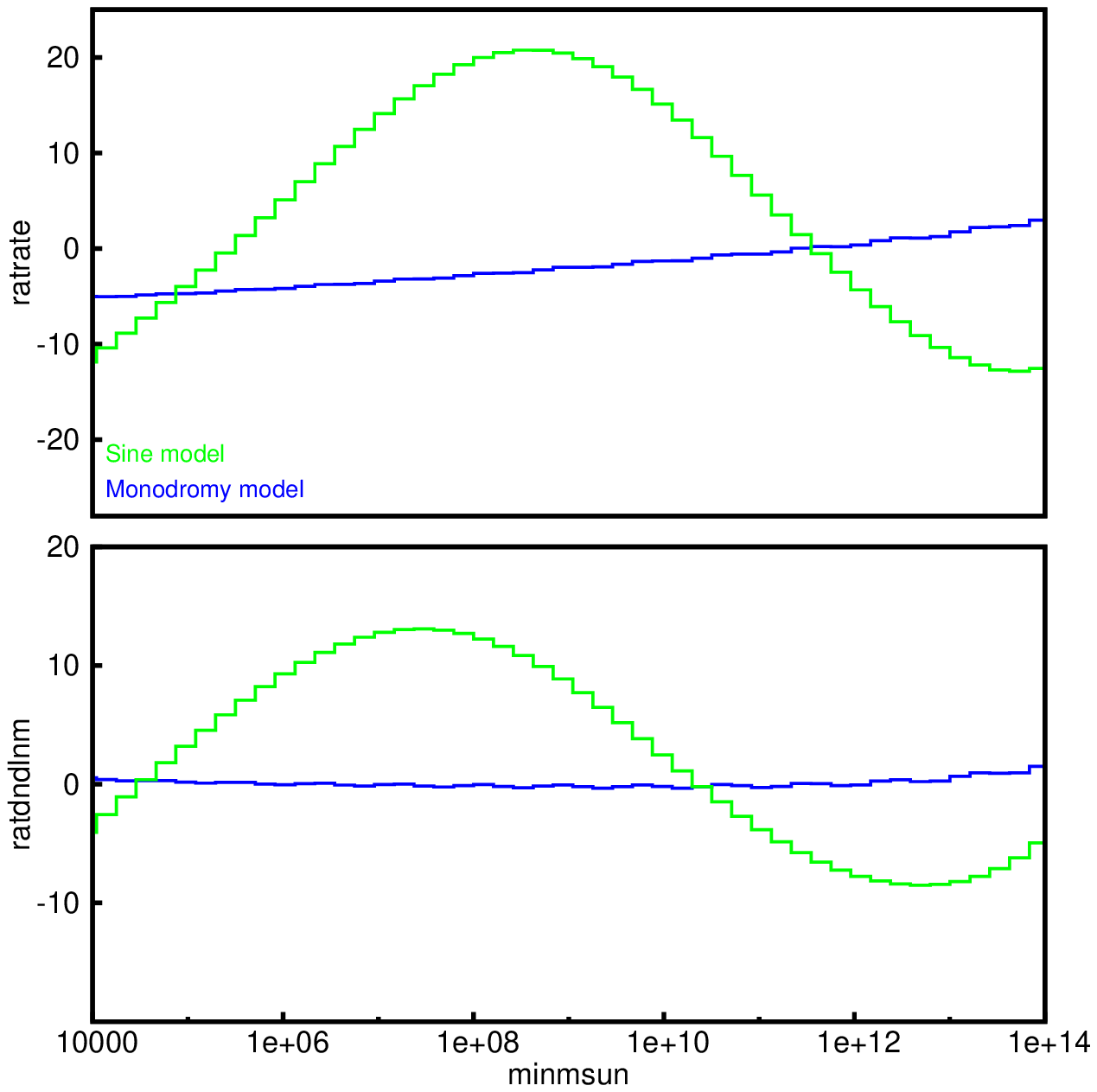}} 
\end{center}
\caption{\footnotesize\label{fig:nh-fbfv} The percentage change in the 
formation rate of halos (in the Press-Schechter formalism, on top) and 
in their number density (in the Sheth-Tormen formalism, at the bottom) 
for the two inflationary models containing oscillatory terms, with respect
to the more conventional quadratic potential. 
The figures on the left correspond to the best fit values 
(indicated by the red points in Figure~\ref{fig:likelihoods}), while those 
on the right correspond to values chosen within (in fact, {\it on})\/ 
the $2$-$\sigma$ confidence contours of the parameters $\alpha$ and 
$\beta$, determined from the joint constraints of the WMAP and SDSS data 
(indicated by the blue points in Figure~\ref{fig:likelihoods}).
In order to highlight the effects due to the primordial features, we have 
frozen the background parameters at the best fit values arrived at when 
the primordial spectrum is determined by the quadratic potential. 
We have then worked with the best fit values for the potential parameters 
to arrive at the figures on the left. 
We have plotted the percentage change in logarithmic mass bins,  i.e. 
$\Delta\log_{10} (M/M_{\odot})$, of $0.2$. 
It is clear that the features corresponding to the best fit values do not 
lead to any substantial difference in either the number or the formation 
rates of the halos. 
However, note that the quadratic potential with superimposed sinusoidal 
oscillations leads to a $13\%$ change in the number of halos formed when 
we choose to work with values of $\alpha$ and $\beta$ that lie within the 
$2$-$\sigma$ contours.}
\end{figure}

In this subsection, we shall discuss the effects of the features on the 
number density and the formation rates of halos in the different 
inflationary models of our interest. 
In order to highlight the effects purely due to the primordial features, 
we have frozen the values of the background cosmological parameters, viz. 
$\Omega_{\rm m}$, $\Omega_{\Lambda}$, $H_0$ and the dimensionless 
baryon density parameter $\Omega_{\rm b}$ at the values arrived at upon 
comparing the smooth quadratic potential with the WMAP and SDSS data, as 
listed in Table~\ref{tab:bestfit}.
But, we have made use of the best fit values for the potential parameters
to compute the inflationary scalar power spectrum and from thereon the
matter power spectrum and the number density of halos. 
In Figure~\ref{fig:nh-fbfv}, we have plotted the percentage of change in the 
formation rate of halos in the Press-Schechter formalism and the number density 
of halos in the Sheth-Tormen formalism for different models with respect to the 
quadratic potential. 
In the case of the model with the step, the change in the number density due 
to the step (corresponding to parameter values within 2-$\sigma$ around the 
best fit values) occurs at very high mass halos ($\sim 10^{17} M_{\odot}$) and 
hence lies outside our region of interest. 
Due to this reason, we have only presented the results in the case of the models 
with oscillatory terms in the potential.

\par

In order to arrive at the maximum possible change in the number density of 
halos when compared to the conventional nearly scale invariant primordial 
spectrum, for the models with oscillations in the potential, we have chosen 
values for the parameters $\alpha$ and $\beta$ that they lie within 
(actually, {\it at})\/ $2$-$\sigma$ from the best fit values. 
We have chosen the parameters in such a way that they create the largest 
deviation from the nearly scale invariant power spectra
that are allowed by the CMB and observations of the large scale structure.
In Figure~\ref{fig:nh-fbfv}, apart from the results for the best fit values, 
we have plotted the number density and the formation rates of halos for the 
cases wherein [$\alpha, \ln\, (\beta/\Mpl)$] is set to ($1.7\times10^{-3},-1.5$) 
and ($2\times10^{-4},-3.5$) (as indicated by the blue points in 
Figure~\ref{fig:likelihoods}) for the quadratic potential with sinusoidal 
modulations and the axion monodromy model, respectively. 
In arriving at these plots, we have fixed the parameters $m$ and $\lambda$ 
at their best fit values as shown in the Table~\ref{tab:bestfit}, since 
these parameters do not play a role in altering the features in the spectrum. 
We have also chosen the value of $\delta$ to be the best fit value for both 
the models.
It is evident from the figure that, for the best fit values of the parameters, 
the change in the number density is completely negligible ($\sim 2\%$). 
However, we find that, for the case of the quadratic potential with sinusoidal 
modulation, the numbers can change by as much as $22\%$ for values of the 
potential parameters $\alpha$ and $\beta$ that lie within $2$-$\sigma$. 
It should also be highlighted that the monodromy model does not seem to lead 
to the same extent of change in the number density and the rate of formation 
of halos, despite the fact that it produces fine oscillations in the 
primordial as well as the linear matter power spectra 
(cf. Figures~\ref{fig:sps-all} and~\ref{fig:mps-all}). 
Actually, while the {\it unbinned}\/ number density does indicate a $5$--$15\%$ 
change, we find that, the change proves to be smaller when we bin the numbers 
in logarithmic mass bins, i.e. $\Delta\log_{10} (M/M_{\odot})$, of $0.2$. 
Evidently, binning seems to average out the rapid oscillations, resulting in a 
smaller extent of change in the numbers.


\section{Discussion}\label{sec:discussion}

In this work, we have investigated the effects of primordial features on
the matter power spectrum as well as the number of halos formed and their 
rate of formation. 
Similar work in this context~\cite{rodrigues-2010} had suggested that a 
small change in the parameters describing the inflaton potential would
lead to a drastic change in the number of halos formed. 
The earlier work was based on the inflationary perturbation spectrum that 
was arrived at based on the slow roll approximation.
In contrast, we have carried a complete and accurate numerical analysis.
Further, we have made use of the Sheth-Tormen mass function (instead of
the older Press-Schechter one) which is known to fit the data from the
$N$-body simulations better. We have included the baryon acoustic 
oscillations in our analysis to have a more realistic comparison. 
Moreover, to arrive at the parameter space of interest, all the potentials 
considered in our work have been constrained by an MCMC analysis (using 
COSMOMC) against the WMAP-7 and SDSS LRG DR7 datasets.
We find that, the best fit values for the potential parameters (with the 
background parameters kept fixed) lead to hardly any change in the number 
of halos formed when compared to the conventional quadratic potential that 
generates a nearly scale invariant primordial spectrum. 
However, partly consistent with the earlier result, we find that values for 
the potential parameters that lie within $2$-$\sigma$ of the best fit values 
indeed lead to a reasonable change in the number of halos formed and in their 
formation rates. 
For instance, we find that, with superimposed sinusoidal modulations, the 
quadratic potential leads to as much as a $13$--$22\%$ change in the halo
number density and the rate of formation. 
Needless to mention, the step of comparing the models against the data is 
crucial as this imposes real bounds on the extent of changes in the numbers 
involved. 
It is worthwhile to note that the inclusion of SDSS data reduces the maximum 
change in number density to about $10\%$, when compared to the case wherein 
one works with the parameters constrained by the WMAP data alone.

\par

We would like to close this paper with the following remarks. 
As we had pointed out before, while comparing with the SDSS data, we have not 
taken into account the non-linear effects on the matter power spectrum. 
It is for this reason that the theoretical curve had not fit the observational 
data well on small scales (cf. Figure~\ref{fig:mps-all}). 
Clearly, a more complete analysis should involve modeling of the non-linear 
effects and their inclusion in evaluating the matter power 
spectrum~\cite{nonlinear}. 
For instance, it will be interesting to compare the results on the number of 
halos formed in numerical simulations, evolved from primordial spectra with 
features, with the small scale data.


\section*{Acknowledgments}

The author wishes to thank Shiv Sethi and L.~Sriramkumar for discussions as 
well as comments on the manuscript. Computational work for this study has 
been carried out using the cluster computing facilities at Harish-Chandra 
Research Institute, Allahabad, India (http://cluster.hri.res.in/).



\begin{thebibliography}{99}
\bibitem{texts}
E.~W.~Kolb and M.~S.~Turner, {\sl The Early Universe}\/ (Addison-Wesley, 
Redwood City, California, 1990); S.~Dodelson, {\sl Modern Cosmology}\/ 
(Academic Press, San Diego, U.S.A., 2003); V.~F.~Mukhanov, {\sl Physical 
Foundations of Cosmology}\/ (Cambridge University Press, Cambridge, England, 
2005); S.~Weinberg, {\sl Cosmology}\/ (Oxford University Press, Oxford, 
England, 2008); R.~Durrer, {\sl The Cosmic Microwave Background}\/ (Cambridge
University Press, Cambridge, England, 2008); D.~H.~Lyth and A.~R.~Liddle, 
{\sl The Primordial Density Perturbation}\/ (Cambridge University Press, 
Cambridge, England, 2009); P.~Peter, J-P.~Uzan and J.~Brujic, {\sl Primordial
Cosmology}\/ (Oxford University Press, Oxford, England, 2009).
\bibitem{reviews}
H.~Kodama and M.~Sasaki, Prog.\ Theor.\ Phys.\ Suppl.\ {\bf 78}, 
1 (1984); V.~F.~Mukhanov, H.~A.~Feldman and R.~H.~Brandenberger, 
Phys.\ Rep.\ {\bf 215}, 203 (1992); J.~E.~Lidsey, A.~Liddle, 
E.~W.~Kolb, E.~J.~Copeland, T.~Barreiro and M.~Abney, Rev.\ Mod.\ 
Phys.\ {\bf 69}, 373 (1997); D.~H.~Lyth and A.~Riotto, Phys.\ Rep.\ 
{\bf 314}, 1 (1999);  A.~Riotto, arXiv:hep-ph/0210162; J.~Martin,
arXiv:hep-th/0406011; B.~Bassett, S.~Tsujikawa and D.~Wands, Rev.\ 
Mod.\ Phys.\ {\bf 78}, 537 (2006); W.~H. Kinney, arXiv:0902.1529 
[astro-ph.CO]; L.~Sriramkumar, Curr.\ Sci.\ {\bf 97}, 868 (2009);
D.~Baumann, arXiv:0907.5424v1 [hep-th].
\bibitem{wmap-5}
J.~Dunkley {\it et al.},\/ Astrophys.\ J.\ Suppl.\ {\bf 180}, 306 (2009);
E.~Komatsu {\it et al.},\/ Astrophys.\ J.\ Suppl.\ {\bf 180}, 330 (2009).
\bibitem{wmap-7}
D.~Larson {\it et al.},\/ Astrophys.\ J.\ Suppl.\ {\bf 192}, 16 (2011);
E.~Komatsu {\it et al.},\/ Astrophys.\ J.\ Suppl.\ {\bf 192}, 18 (2011).
\bibitem{act-2010}
J.~Dunkley {\it et al.},\/ Astrophys.\ J.\ {\bf 739}, 52 (2011).
\bibitem{l2240}
J.~A.~Adams, B.~Cresswell and R.~Easther, Phys.\ Rev.\ D\ {\bf 64}, 
123514 (2001); L.~Covi, J.~Hamann, A.~Melchiorri, A.~Slosar and 
I.~Sorbera, Phys.\ Rev.\ D\ {\bf 74}, 083509 (2006); J.~Hamann, 
L.~Covi, A.~Melchiorri and A.~Slosar, Phys.\ Rev.\ D\ {\bf 76}, 
023503 (2007); M.~J.~Mortonson, C.~Dvorkin, H.~V.~Peiris and W.~Hu, 
Phys.\ Rev.\ D\ {\bf 79}, 103519 (2009).
\bibitem{joy-2008-2009}
M.~Joy, V.~Sahni and A.~A.~Starobinsky, Phys.\ Rev.\ D\ {\bf 77}, 023514 
(2008); M.~Joy, A.~Shafieloo, V.~Sahni and A.~A.~Starobinsky, JCAP 
{\bf 0906}, 028 (2009).
\bibitem{pi}
R.~K.~Jain, P.~Chingangbam, J.-O.~Gong, L.~Sriramkumar and T.~Souradeep, 
JCAP {\bf 0901}, 009 (2009); R.~K.~Jain, P.~Chingangbam, L.~Sriramkumar 
and T.~Souradeep, Phys.\ Rev.\ D\ {\bf 82}, 023509 (2010). 
\bibitem{hazra-2010}
D.~K.~Hazra, M.~Aich, R.~K.~Jain, L.~Sriramkumar and T.~Souradeep, JCAP 
{\bf 1010}, 008 (2010).
\bibitem{starobinsky-1992}
A.~A.~Starobinsky, Sov.\ Phys.\ JETP\ Lett.\ {\bf 55}, 489 (1992).
\bibitem{hu-2010-2011}
C.~Dvorkin and W.~Hu, Phys.\ Rev.\ D\ {\bf 81}, 023518 (2010); W.~Hu, 
Phys.\ Rev.\ D\ {\bf 84}, 027303 (2011).
\bibitem{rc}
S.~Hannestad, Phys.\ Rev.\ D\ {\bf 63}, 043009 (2001); S.~L.~Bridle, 
A.~M.~Lewis, J.~Weller and G.~Efstathiou, Mon.\ Not.\ Roy.\ Astron.\ 
Soc.\  {\bf 342}, L72 (2003); P.~Mukherjee and Y.~Wang, Astrophys.\ 
J.\ {\bf 599}, 1 (2003); S.~Hannestad, JCAP\ {\bf 0404}, 002 (2004); 
A.~Shafieloo and T.~Souradeep, Phys.\ Rev.\ D\ {\bf 70}, 043523 (2004);
D.~Tocchini-Valentini, Y.~Hoffman and J.~Silk, Mon.\ Not.\ Roy.\ 
Astron.\ Soc.\ {\bf 367}, 1095 (2006); A.~Shafieloo, T.~Souradeep, 
P.~Manimaran, P.~K.~Panigrahi and R.~Rangarajan, Phys.\ Rev.\ D\ {\bf 75}, 
123502 (2007); A.~Shafieloo and T.~Souradeep, Phys.\ Rev.\ D\ {\bf 78}, 
023511 (2008); R.~Nagata and J.~Yokoyama, Phys.\ Rev.\ D\ {\bf 79}, 043010 
(2009); G.~Nicholson and C.~R.~Contaldi, JCAP\ {\bf 0907}, 011 (2009).
\bibitem{pahud-2009}
C.~Pahud, M.~Kamionkowski and A.~R.~Liddle, Phys.\ Rev.\ D\ {\bf 79}, 083503 
(2009).
\bibitem{flauger-2010}
R.~Flauger, L.~McAllister, E.~Pajer, A.~Westphal and G.~Xu, JCAP {\bf 1006}, 
009 (2010).
\bibitem{aich-2011}
M.~Aich, D.~K.~Hazra, L.~Sriramkumar and T.~Souradeep, arXiv:1106.2798v2
[astro-ph.CO].
\bibitem{ng}
J.~Maldacena, JHEP\ {\bf 0305}, 013 (2003); D.~Seery and J.~E.~Lidsey, JCAP 
{\bf 0506}, 003 (2005); X.~Chen, Adv.\ Astron.\ {\bf 2010}, 638979 (2010).
\bibitem{ng-features}
X.~Chen, R.~Easther, E.~A.~Lim, JCAP {\bf 0804}, 010 (2008); R.~Flauger 
and E.~Pajer, JCAP {\bf 1101}, 017 (2011); F.~Arroja, A.~E.~Romano and 
M.~Sasaki, Phys.\ Rev.\ D\ {\bf 84}, 123503 (2011); J.~Martin and 
L.~Sriramkumar, JCAP {\bf 1201}, 008 (2012); D.~K.~Hazra, L.~Sriramkumar 
and J.~Martin, arXiv:1201.0926v1 [astro-ph.CO]; F.~Arroja and M.~Sasaki, 
JCAP {\bf 1208}, 012 (2012). 
\bibitem{rodrigues-2010}
L.~F.~S.~Rodrigues and R.~Opher, Phys.\ Rev.\ D\ {\bf 82}, 023501 (2010).
\bibitem{sdss}
See, for instance, {\tt http://www.sdss.org/}
\bibitem{sdss-data}
K.~Abazajian {\it et. al.} Astrophys.\ J. \ Suppl. {\bf 182}, 543-558 (2009)
\bibitem{jain-2007}
R.~K.~Jain, P.~Chingangbam and L. Sriramkumar, JCAP 10, 003 (2007).
\bibitem{camb}
See, {\tt http://camb.info/}.
\bibitem{lewis-2000}
A.~Lewis, A.~Challinor and A.~Lasenby, Astrophys.\ J.\ {\bf 538}, 473 (2000).
\bibitem{cosmomc}
See, {\tt http://cosmologist.info/cosmomc/}.
\bibitem{lewis-2002}
A.~Lewis and S.~Bridle, Phys.\ Rev.\ D\ {\bf 66}, 103511 (2002).
\bibitem{huang}
Z.~Huang, JCAP {\bf 1206}, 012 (2012).
\bibitem{nonlinear}
R.~E.~Smith {\it et al.},\/ Mon.\ Not.\ Roy.\ Astron.\ Soc.\  {\bf 341}, 
1311 (2003); M.~Tegmark {\it et al.},\/ Phys.\ Rev.\ D {\bf 74}, 123507 
(2006); W.~J.~Percival {\it et al.},\/ Astrophys.\ J.\  {\bf 657}, 645 
(2007).
\bibitem{mo-2010}
H.~Mo, F.~v.~d.~Bosch and S.~White, {\sl Galaxy Formation and Evolution}\/
(Cambridge University Press, Cambridge, England, 2010).
\bibitem{takada-2006}
M.~Takada, E.~Komatsu and T.~Futamase, Phys.\ Rev.\ D\ {\bf 73}, 083520 (2006).
\bibitem{gf}
L.~Wang and P.~J.~Steinhardt, Astrophys.\ J.\ {\bf 508}, 483 (1998); 
E.~V.~Linder and A.~Jenkins, Mon.\ Not.\ Roy.\ Astron.\ Soc.\ {\bf 346}, 
573 (2003).
\bibitem{jenkins-2001}
A.~Jenkins, C.~S.~Frenk, S.~D.~M.~White, J.~M.~Colberg, S.~Cole, A.~E.~Evrard,
H.~M.~P.~Couchman and N.~Yoshida, Mon.\ Not.\ Roy.\ Astron.\ Soc.\ {\bf 321}, 
372 (2001).
\bibitem{sheth-2001-2002}
R.~K.~Sheth, H.~J.~Mo and G.~Tormen, Mon.\ Not.\ Roy.\ Astron.\ Soc.\ 
{\bf 323}, 1 (2001); R.~K.~Sheth and G.~Tormen, Mon.\ Not.\ Roy.\ Astron.\ 
Soc.\  {\bf 329}, 61 (2002).
\bibitem{press-1974}
W.~H.~Press and P.~Schechter, Astrophys.\ J.\ {\bf 187}, 425 (1974).
\bibitem{hfr}
S.~Sasaki, Publ.\ Astron.\ Soc.\ Jap. {\bf 46}, 427 (1994); E.~Ripamonti, 
Mon.\ Not.\ Roy.\ Astron.\ Soc.\ {\bf 376}, 709 (2007); S.~Mitra, G.~Kulkarni, 
J.~S.~Bagla and J.~K.~Yadav, Bull.\ Astron.\ Soc.\ Ind.\ {\bf 39}, 1,(2011).
\bibitem{press-1996}
W.~H.~Press, S.~A.~Teukolsky, W.~T.~Vetterling and B.~P.~Flannery, 
{\sl Numerical Recipes in Fortran~90},\/ Second edition (Cambridge 
University Press, Cambridge, England, 1996).
\bibitem{komatsu-code}
See, {\tt http://www.mpa-garching.mpg.de/$\sim$komatsu/CRL/index.html}.
\bibitem{gq}
See, for instance, {\tt http://www.nag.com/} and {\tt http://www.netlib.org/}.
\bibitem{finelli-2010}
F.~Finelli, J.~Hamann, S.~M.~Leach and J.~Lesgourgues, JCAP {\bf 04} 011 (2010).
\bibitem{benetti-2012}
M.~Benetti, S.~Pandolfi, M.~Lattanzi, M.~Martinelli and A.~Melchiorri,
arXiv:1210.3562 [astro-ph.CO].
\end{thebibliography}
\end{document}